\documentclass[preprint,showpacs,preprintnumbers,amsmath,amssymb,aps,nofootinbib]{revtex4}

\newcommand{\tmpand}{}
\newcommand{\tmpred}{}
\newcommand{\tmpblue}{}
\newcommand{\tmpgreen}{}
\newcommand{\tmpcyan}{}
\newcommand{\tmpmagenta}{}

\usepackage{graphicx}
\usepackage{dcolumn}
\usepackage{bm}

\def\LP{\left(}		
\def\RP{\right)}	

\newcommand{\BE}{\begin{equation}}
\def\EE{\end{equation}}
\def\BEA{\begin{eqnarray}}
\def\EEA{\end{eqnarray}}
\def\EL{\nonumber\\}

\begin{document}

\title{Lattice calculation of $1^{-+}$ hybrid mesons with improved
Kogut-Susskind fermions}

\author{C. Bernard} 
\affiliation{Department of Physics, Washington University,
St.~Louis, MO 63130, USA}

\author{T. Burch} 
\affiliation{Department of Physics, University of 
Arizona, Tucson, AZ 85721, USA}

\author{C. DeTar} 
\affiliation{Physics Department, University of Utah, 
Salt Lake City, UT 84112, USA}

\author{Steven Gottlieb} 
\affiliation{Department of Physics, Indiana 
University, Bloomington, IN 47405, USA}

\author{E.B. Gregory} 
\affiliation{Department of Physics, University of 
Arizona, Tucson, AZ 85721, USA}

\author{U.M. Heller} 
\affiliation{American Physical Society, One Research Road,
Box 9000, Ridge, NY 11961-9000}

\author{J. Osborn} 
\affiliation{Physics Department, University of Utah, 
Salt Lake City, UT 84112, USA}

\author{R. Sugar}
\affiliation{Department of Physics, University of California, 
Santa Barbara, CA 93106, USA}

\author{D. Toussaint} 
\affiliation{Department of Physics, University of 
Arizona, Tucson, AZ 85721, USA}

\date{\today}
    
\begin{abstract}
We report on a lattice determination of the mass of the exotic $1^{-+}$ hybrid
meson using an improved Kogut-Susskind action. Results from
both quenched and dynamical quark simulations are presented.
We also compare with earlier results using Wilson quarks at heavier quark
masses.
The results on lattices with three flavors of dynamical quarks show
effects of sea quarks on the hybrid propagators which probably result
from coupling to two meson states.
We extrapolate the quenched results to the physical light quark mass
to allow comparison with experimental candidates for the $1^{-+}$ hybrid meson.
The lattice result remains somewhat heavier than the experimental result,
although
it may be consistent with the $\pi_1(1600)$.
\end{abstract}
   

\maketitle

\section{INTRODUCTION}
The fact that gluons carry color charge suggests that they, like quarks,
could be ``valence'' constituents of hadrons.  In other words, we expect
that the spectrum of QCD should contain glueballs and hybrids, or particles
with both quarks and gluons as valence constituents.
Hybrid mesons can have exotic quantum numbers, or $J^{PC}$ combinations
not possible with a quark-antiquark state.
However, a state with exotic quantum numbers is not necessarily a
hybrid --- it could be a $\bar q \bar q q q$ state, realized either
as a single ``bag'' containing four quarks or as a ``molecule''
made of two $\bar q q$ mesons.
Experimental evidence suggests the existence of one or more mesons with exotic 
quantum numbers $J^{PC}=1^{-+}$, namely the $\pi_1(1400)$ ~\cite{pi1400} 
and the $\pi_1(1600)$ ~\cite{pi1600}.
Analytic and numerical methods to predict the mass of light 
hybrid meson states include
flux tube models ~\cite{Barnes:1995hc}, 
the bag model~\cite{Flensburg,Barnes_bag,Chanowitz,Barnes_bag2,Barnes_bag3}, 
QCD spectral sum rules~\cite{Chetyrkin:2000tj,Latorre:1985tg,Narison:1999hg},
relativistic WKB calculations~\cite{Cornwall}, and lattice QCD.
Several lattice studies ~\cite{Lacock:1996ny,Bernard:1997ib,ZSU} 
have used quenched Wilson or quenched Wilson-clover fermions 
to calculate the masses of exotic hybrid states, although with
quark masses much larger than the physical $u$ and $d$ quark masses.
Lacock and Schilling have done a calculation in two flavor QCD,
again with fairly heavy quarks~\cite{Lacock_dyn}.

Here we report results of a lattice calculation of the mass of a $1^{-+}$ hybrid
meson using improved Kogut-Susskind quarks.  The use of Kogut-Susskind
quarks allows us to work at valence quark masses much smaller than were used
in previous lattice calculations.  In addition, the ``$a^2_{\rm tad}$''
action that we use has leading lattice spacing errors of order $a^2 g^2$,
while the clover-Wilson action has errors of order $a^2$.
Our mass estimates in the quenched approximation are consistent with
earlier Wilson quark results, but extrapolation to the physical light
valence quark masses is under much better control.
Preliminary results of this calculation were reported in Ref.~\cite{LAT02_hybrids}.

We have also calculated hybrid meson propagators including the effects
of three flavors of dynamical quarks, with light sea quark masses down
to $0.4$ times the strange quark mass.  We find that extracting mass
estimates from the propagators in full QCD is difficult, and we
argue that this difficulty is due to mixing of the hybrid meson
with two meson states --- the states into which it might decay.

\section{$1^{-+}$ HYBRID MESON OPERATOR}

We can construct a $1^{-+}$ hybrid meson operator as the cross product of 
a color octet $1^{--}$ quark-antiquark ($\rho$ meson) operator and the chromomagnetic field,
which has $J^{PC}=1^{+-}$: 
$\rho \times B$ ~\cite{Bernard:1997ib}.
With staggered quarks we have several 
choices of rho meson operators, but it is convenient to 
choose the taste\ \footnote{
We use the term ``taste'' to refer to the four types of quarks that
are naturally present in the Kogut-Susskind formulation, while
``flavor'' can also distinguish quarks with an additional externally
imposed label.  For example, a meson with a source operator
$\bar\psi \gamma_5 \otimes {\bf 1} \psi$ but with disconnected diagrams
not included would be a taste singlet but flavor non-singlet, and would
be a pion in the continuum limit.}
singlet $\rho_s$, with the spin $\otimes$ taste structure
$\gamma_i \otimes {\mathbf 1}$. 
\BEA
1^{-+}_i &=& \epsilon_{ijk}\, \bar\psi^a\, \gamma_j \otimes {\bf 1}\, \psi^b\, B_k^{ab} \EL
&=& 2\, \bar\psi^a\, \gamma_j \otimes {\bf 1}\, \psi^b\, F_{ij}^{ab} \ \ \ ,
\EEA
where $i$, $j$ and $k$ are spatial indices and $a$ and $b$ are color indices.
Each spin component of the $1^{-+}$ includes two terms, for 
example:
\begin{equation}
1^{-+}_x = \rho_y B_z - \rho_z B_y\ \ \ ,
\end{equation}
so if we had chosen a spin $\otimes$ taste structure like
$\gamma_i \otimes \gamma_i$ the two components of $1^{-+}_x$ would
have different tastes.

The Kogut-Susskind $\rho_s$ meson operator, with spin aligned in the $k$ 
direction is
$\bar{\chi}\eta_k D_k \chi,$ ~\cite{Golterman:1985dz}
where $\chi$ and $\bar{\chi}$ are the quark and antiquark fields respectively.
The covariant symmetric shift operator is given by
\begin{equation}
D_{\mu}q(x) = \frac{1}{2} \left[ U_{\mu}^\dagger(x-\hat \mu)q(x-\hat\mu)+U_{\mu}(x)q(x+\hat\mu) \right] \ \ \ .
\end{equation}

We compute the field strength at each lattice point using the
four plaquettes in each plane that have corners at this point,
as described in Ref.~\cite{Bernard:1997ib}.
In computing the field strength, we use links that have
been smoothed with 32
iterations of APE smearing in the spatial directions only
with relative weight of the staples set to 0.25 ~\cite{APE}.
This smearing removes short wavelength fluctuations in the gluon
field, and reduces the noise in the hybrid propagator.
(The smeared links are only used in constructing $F_{\mu\nu}$; the propagators
are computed using the original links.)

\begin{figure}[t]
\resizebox{3.0in}{!}{\includegraphics{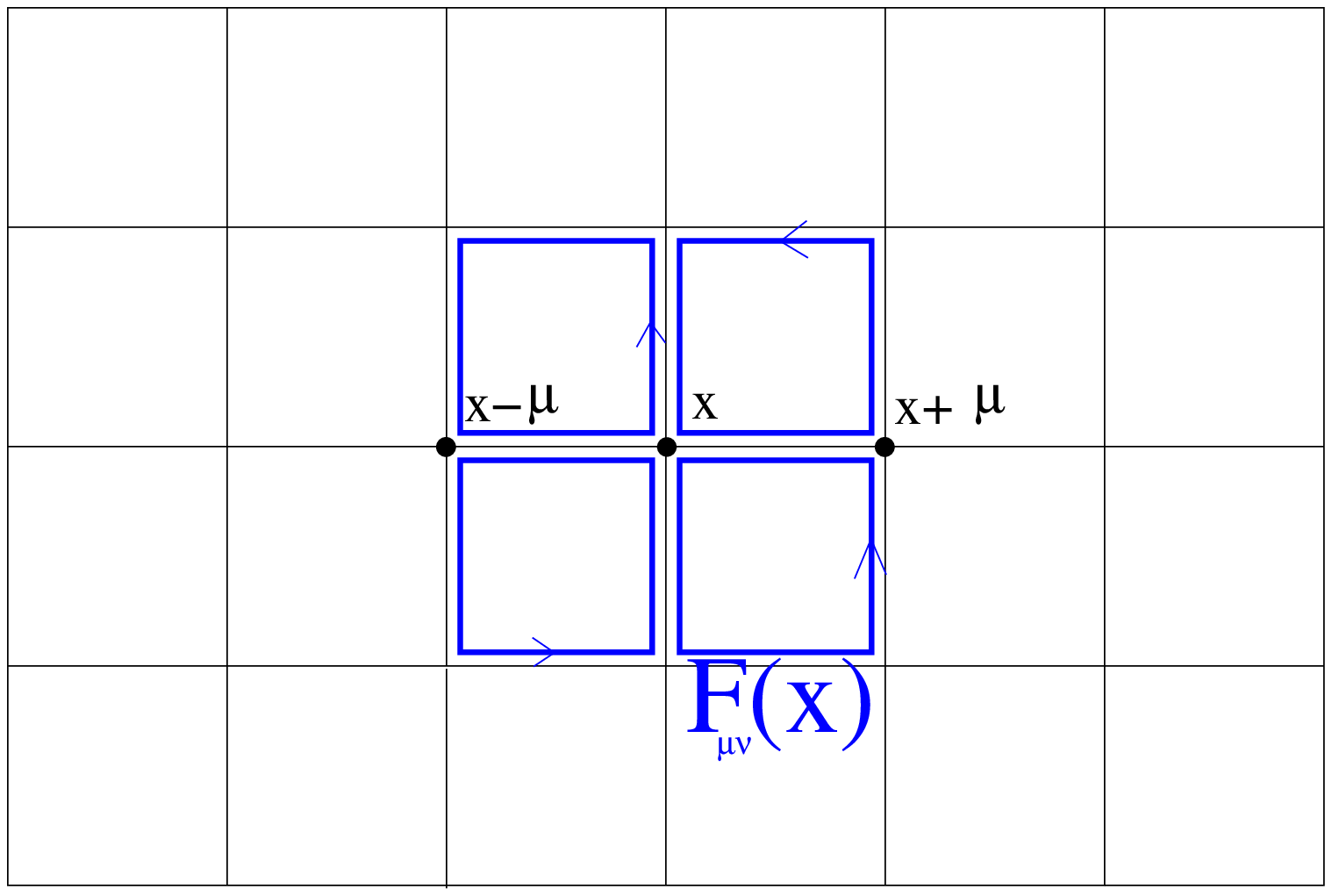}}
\resizebox{3.0in}{!}{\includegraphics{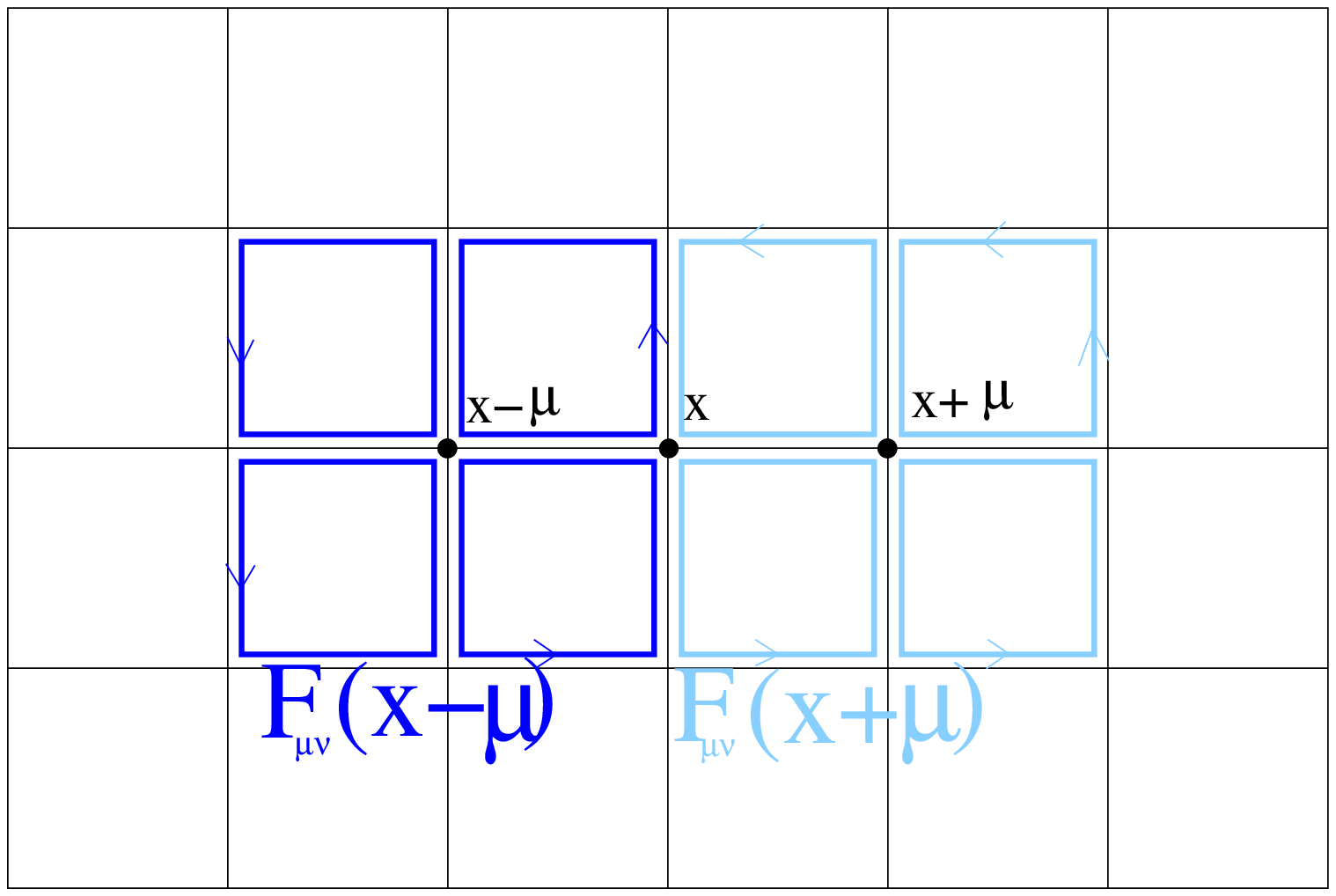}}
\caption{\label{aquark_field} Chromomagnetic field measured at the site of 
the antiquark (left) and the quark (right). }
\end{figure}

Our zero momentum hybrid source and sink wave functions are
constructed in Coulomb gauge and consist of a product of quark and
antiquark fields with phases and offsets appropriate to a color octet
$\rho_s$, as described above, and multiplied by the smeared field strength
symmetrized with respect to the positions of the quark or antiquark
to form the required $C$ even
combination, as illustrated in Fig.~\ref{aquark_field}:
\BE \bar \chi \epsilon_{ijk} \LP \eta_i D_i B_j + 
       B_j \eta_i D_i \RP \chi \EE
The operator is summed over all spatial sites and a trace is taken
over the color indices.

The algorithm for constructing the meson propagator starts in Coulomb
gauge with a quark ``wall source'', consisting of a unit color vector
field in a spatially constant direction, and applies the hybrid meson
operator to form a source for the antiquark propagator.  The
calculation of the meson propagator is completed by acting upon the
resulting antiquark propagator at an arbitrary time slice by the same
hybrid operator and joining the resulting color vector field with the
quark field propagated from the same wall source, summing over all
sink spatial sites and color indices.  The whole process is repeated,
summing over the three wall source colors.

\section{SIMULATION AND MEASUREMENT}
We measured the connected correlator of the $1^{-+}$ hybrid state on three 
sets of $ 28^3 \times 96 $ lattices generated with 
the ``$a^2_{\rm tad}$'' action ~\cite{Asqtad}. To isolate the effects of 
dynamical quarks, we used matched quenched and full QCD lattices with
$10/g^2=8.40$, $m_{\rm val}a=0.016,0.04$, for the quenched quarks, 
$10/g^2=7.18$ for lattices with 
three degenerate flavors of dynamical sea quarks at the strange quark 
mass ($ma=0.031$) and  $10/g^2=7.11$ for lattices with 
$m_{u,d}=0.4m_s$ ($ma=0.0124$). These
choices of $10/g^2$ give approximately the same lattice spacing 
($\sim 0.09$ fm) in the three cases.  The corresponding choices of quark mass
allow simulation at roughly equivalent values of $(m_{\rm PS}/m_V)^2$, 
the square of the ratio of the pseudoscalar to vector meson masses.
Table \ref{hybrids_table}
summarizes the simulation parameters and fit results for the $1^{-+}$
states, while Table \ref{conventional_table} contains estimates for
conventional hadron masses at these parameters.

The size of the datasets is 
comparable for quenched and full QCD runs.
Successive full QCD lattices are separated by six molecular dynamics
trajectories, with each trajectory one simulation time unit long.
The full QCD lattices are not completely decorrelated but
this autocorrelation has negligible effects on 
the hybrid mass fittings, since hybrid propagators have much larger statistical
errors than, e.g. pion propagators. 
In particular, for the lightest sea and valence quark mass, $a
m_q=0.0124$, we calculated the normalized autocorrelations of the $1^{-+}$
propagators separated by six simulation time
units at each Euclidean time separation, or distance between
the wall source and sink.  For propagation distances zero
through eight with the sample of 532 lattices we find
 0.01, 0.13, -0.05, -0.00, 0.08, 0.01, -0.04, -0.16 and -0.08 respectively,
instead of the uniformly positive autocorrelations that we would
see if the propagators were systematically correlated from one
stored lattice to the next.
Although the statistical errors we quote come from the covariance matrix
of the propagator, we have also performed a jackknife error analysis of each
fitted 
mass and found jackknife error estimates to be consistent with errors from 
the covariance matrix. Varying the block size from 1 to 10 
had no significant effect on the jackknife error.

In a separate study we have measured propagators of the
pion, rho and nucleon.  Statistical errors on these propagators are
much smaller than for the $1^{-+}$ propagator, so some effects of
autocorrelations can be seen.
For the nucleon at mass $am_1=0.0124$, which we use for comparison
with the hybrid propagators, the data was grouped in blocks of
four lattices, or 24 trajectories, before the covariance matrix
was computed.  Further blocking does not significantly increase the
error bars.  The fact that the nucleon mass fits have good $\chi^2$
(in fact, better than the quenched nucleon fits)
is also evidence that this blocking has removed most of the effects
of the autocorrelations.

\section{RESULTS}
We fit the measured correlators to the sum of oscillating and normal 
exponentials:
\begin{equation}
C(t)=A_1e^{-M_{1^{-+}}t}+ A_2(-1)^te^{-m_2t} + A_3(-1)^te^{-m_3t},
\end{equation}
where $M_{1^{-+}}$ is the hybrid meson mass of interest and $m_2$ and $m_3$ are
masses of non-exotic parity partner states which have oscillating correlators in
the Kogut-Susskind formulation.
In our case the oscillating parity partner is a $1^{++}$ ($a_1$) state,
which is lighter than the $1^{-+}$ hybrid, and the oscillating
component dominates the correlator at large times.
It is therefore essential to include the oscillating state(s) in our fits.
We performed both four and five parameter fits. For the four parameter fits,
we fix $A_3=m_3=0$, meaning that we include one state of each parity.
For the five parameter fits we fix $m_2$ to an
$a_1$ meson mass determined from propagators with a standard
$\bar q q$ source operator, and fit for $A_2$, $m_3$ and $A_3$. We varied the 
range of the fit and tried to choose values for $M_{1^{-+}}$ corresponding to 
high-confidence fits that were insensitive to $D_{\rm max}$ 
and $D_{\rm min}$, the limits of the fit range. 

\begin{figure}[t]
\resizebox{3.2in}{!}{\includegraphics{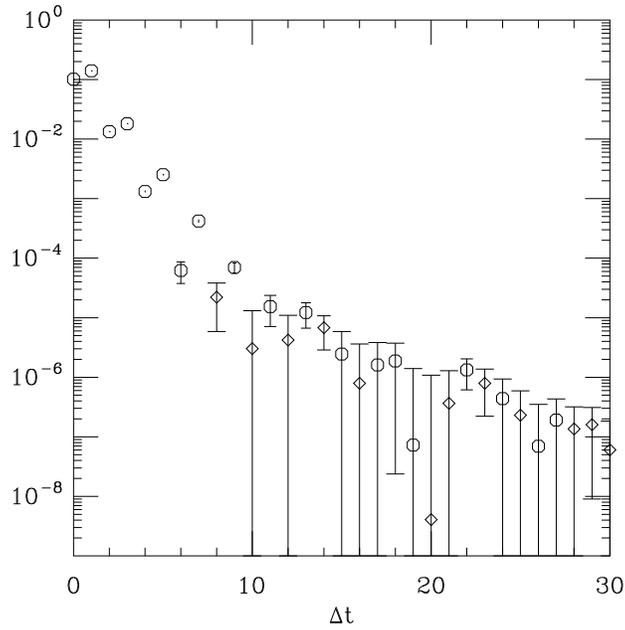}}
\caption{\label{propm016nf0b840}Propagator for quenched lattice with 
$10/g^2=8.40$, ma=0.016. Octagons represent positive values, diamonds represent 
negative values.}
\end{figure}

\begin{figure}[t]
\resizebox{5.0in}{!}{\includegraphics{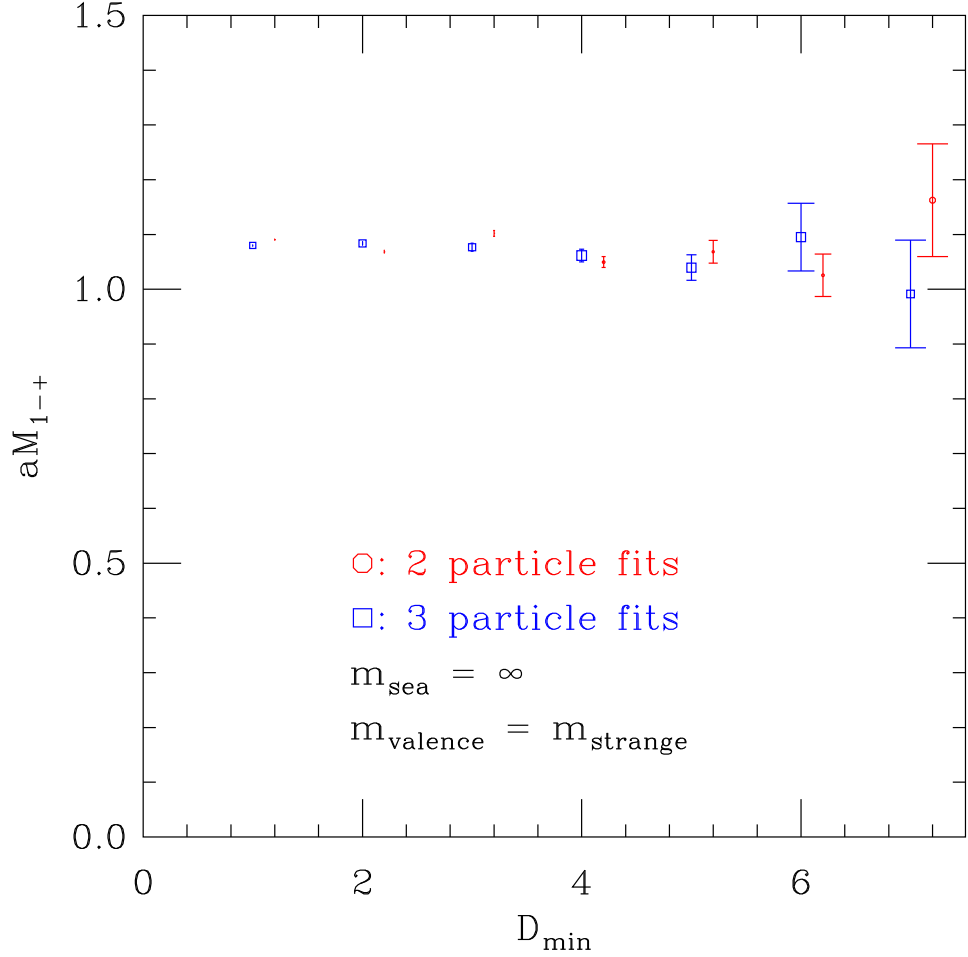}}
\caption{$aM_{1^{-+}}$ {\em vs.} $D_{\rm min}$ for $10/g^2=8.40$ quenched 
lattices $am_{\rm valence}=0.040$.
The \tmpred octagons are four parameter fits, with one mass and
amplitude of each parity, and the \tmpblue squares are five parameter
fits with one $1^{++}$ mass fixed to the $a_1$ mass, as described in the
text.
All these fits used a maximum distance $D_{\rm max}=15$.
The four parameter fit points are shifted slightly to the right
for clarity.
The symbol size is proportional to the confidence level of the fit,
with the symbol size in the labels corresponding to 50\%.
\label{fits_b840m040_fig}}
\end{figure}

\begin{figure}[t]
\resizebox{5.0in}{!}{\includegraphics{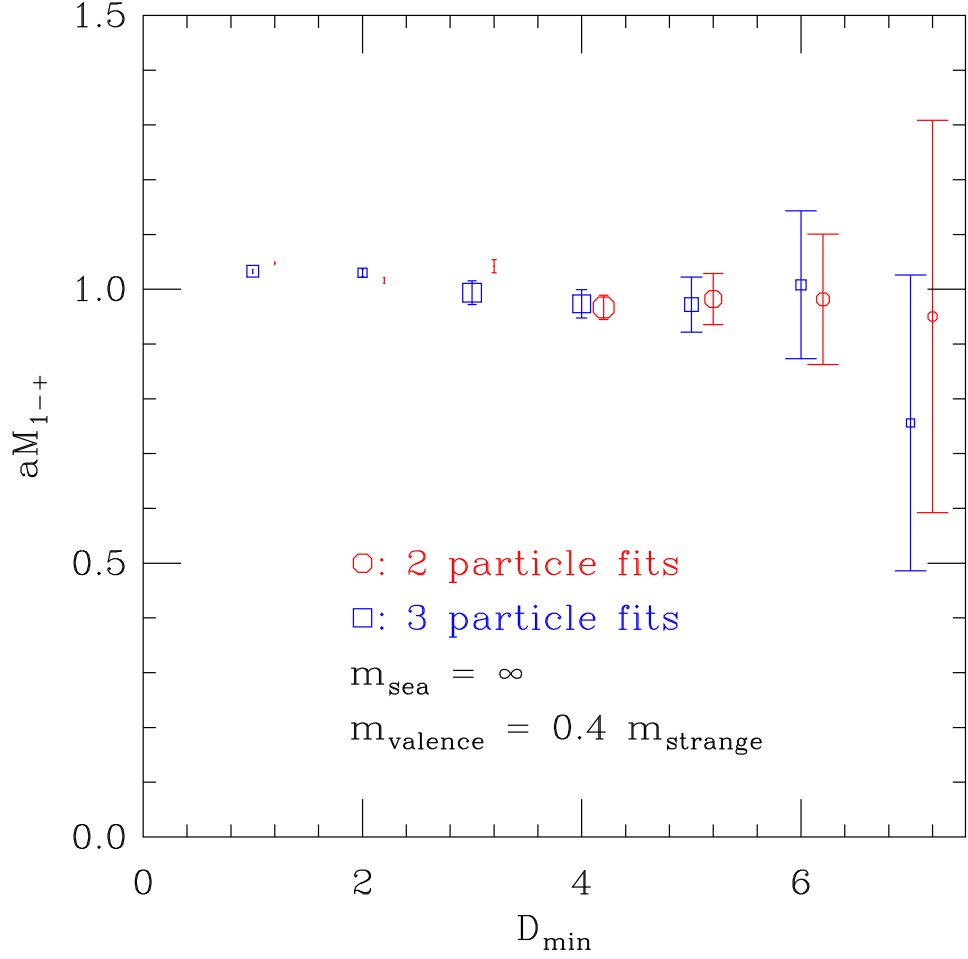}}
\caption{$aM_{1^{-+}}$ {\em vs.} $D_{\rm min}$ for $10/g^2=8.40$ quenched 
lattices with $am_{\rm valence}=0.016$, using $D_{\rm max}=15$.
Notation is the same as in Fig.~\protect\ref{fits_b840m040_fig}.
\label{fits_b840m016_fig}}
\end{figure}

For the quenched lattices we were able to fit the propagators with reasonable 
confidence levels (25-50\%) for valence quark masses $ma=0.016$ and $ma=0.040$.
Figure \ref{propm016nf0b840}  shows the measured 
propagator for $ma=0.016$. Note the oscillating component due
to parity partner states.  As expected, the oscillating component dominates
at large distance, since the parity partner has lower mass than the $1^{-+}$.
Figure~\ref{fits_b840m040_fig}
shows mass fits for the quenched lattices for $ma=0.040$ and 
$D_{\rm max}=15$, with both the two particle (four parameter) and
three particle (five parameter) fits.  In the mass fit plots, we have included 
the small confidence level fits to illustrate how adjusting the fit range produces more optimal fits. 
Figure~\ref{fits_b840m016_fig} shows the same plot for $ma=0.016$.
In both plots the three particle fits exhibit a plateau with relatively small error bars
($<1\%$),
demonstrating the stability of the result with respect to variations in the 
fit range.
For the four-parameter fits, there is a 
slight oscillation of fitted values about the same plateau. Furthermore, 
the range of fits with high confidence level and relatively small errors is
reduced.
From plots like these, we picked a ``best fit'', a value that met
some balance of the following criteria: insensitivity to fit range, high 
confidence level, reasonable statistical errors. We can see that one might 
reasonably choose any one of several points as a ``best fit'', and the
range of resulting $M_{1^{-+}}$ values is the basis of our estimate of 
the systematic error coming from the presence of higher mass
states in the propagators. In all of these fit summary figures we include 
unused fits, that do not meet these criteria, say, because of low confidence 
level, to help illustrate how we selected the optimal fits.

\begin{figure}[t]
\resizebox{5.0in}{!}{\includegraphics{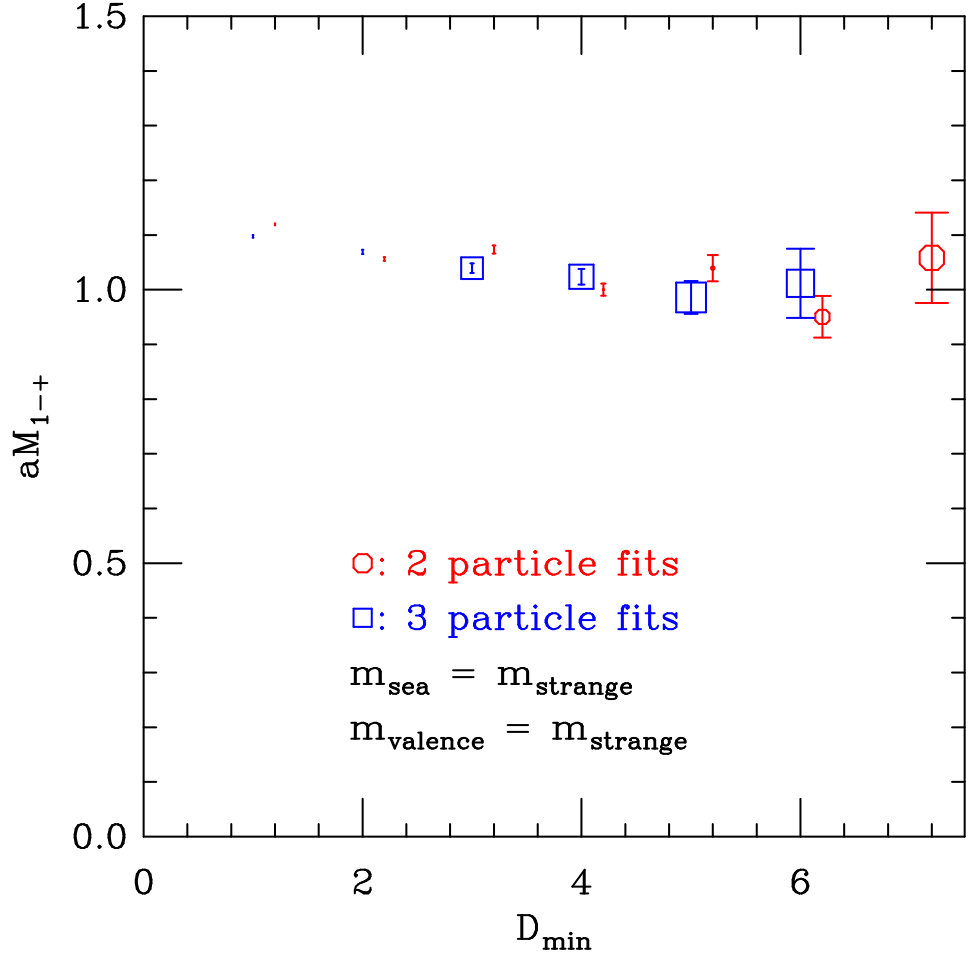}}
\caption{$aM_{1^{-+}}$ {\em vs.} $D_{\rm min}$ for $10/g^2=7.18$ with
three degenerate dynamical quarks with mass $am_{\rm sea}=am_{\rm valence}=0.031$,
using $D_{\rm max}=15$.
Notation is the same as in Fig.~\protect\ref{fits_b840m040_fig}.
\label{fits_b718m031_fig}}
\end{figure}

For lattices with three degenerate sea quarks at $m_s$, we were also able to 
extract a value
for $M_{1^{-+}}$ in reasonable agreement with the quenched result.
Four and 
five parameter fits are shown in Fig. \ref{fits_b718m031_fig}.
The fits exhibit larger statistical errors than the quenched
lattice fits, and a slight dependence on range.
The mass estimate in Table~\ref{hybrids_table}
reflects this with significantly larger 
statistical and systematic error bars than in the quenched case.

\begin{figure}[t]
\resizebox{3.2in}{!}{\includegraphics{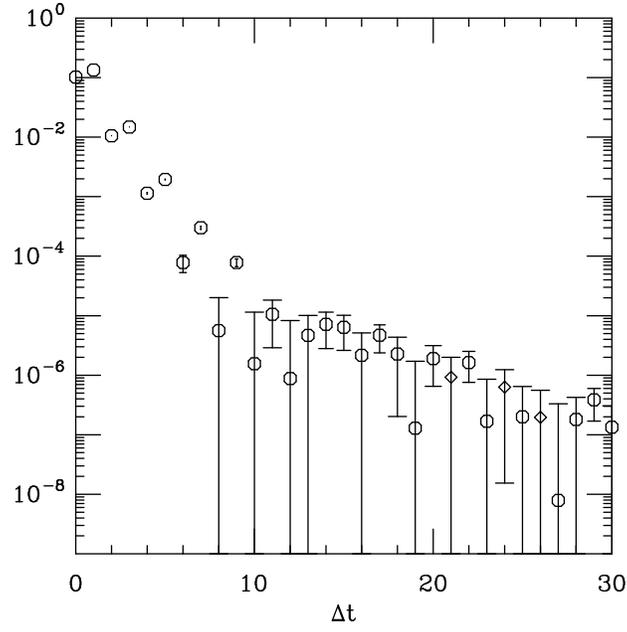}}
\caption{\label{propm0124nf0b711}Propagator for three flavor lattice with 
$10/g^2=7.11$, ma=0.0124. Octagons represent positive values, diamonds represent 
negative values.}
\end{figure}

\begin{figure}[t]
\resizebox{5.0in}{!}{\includegraphics{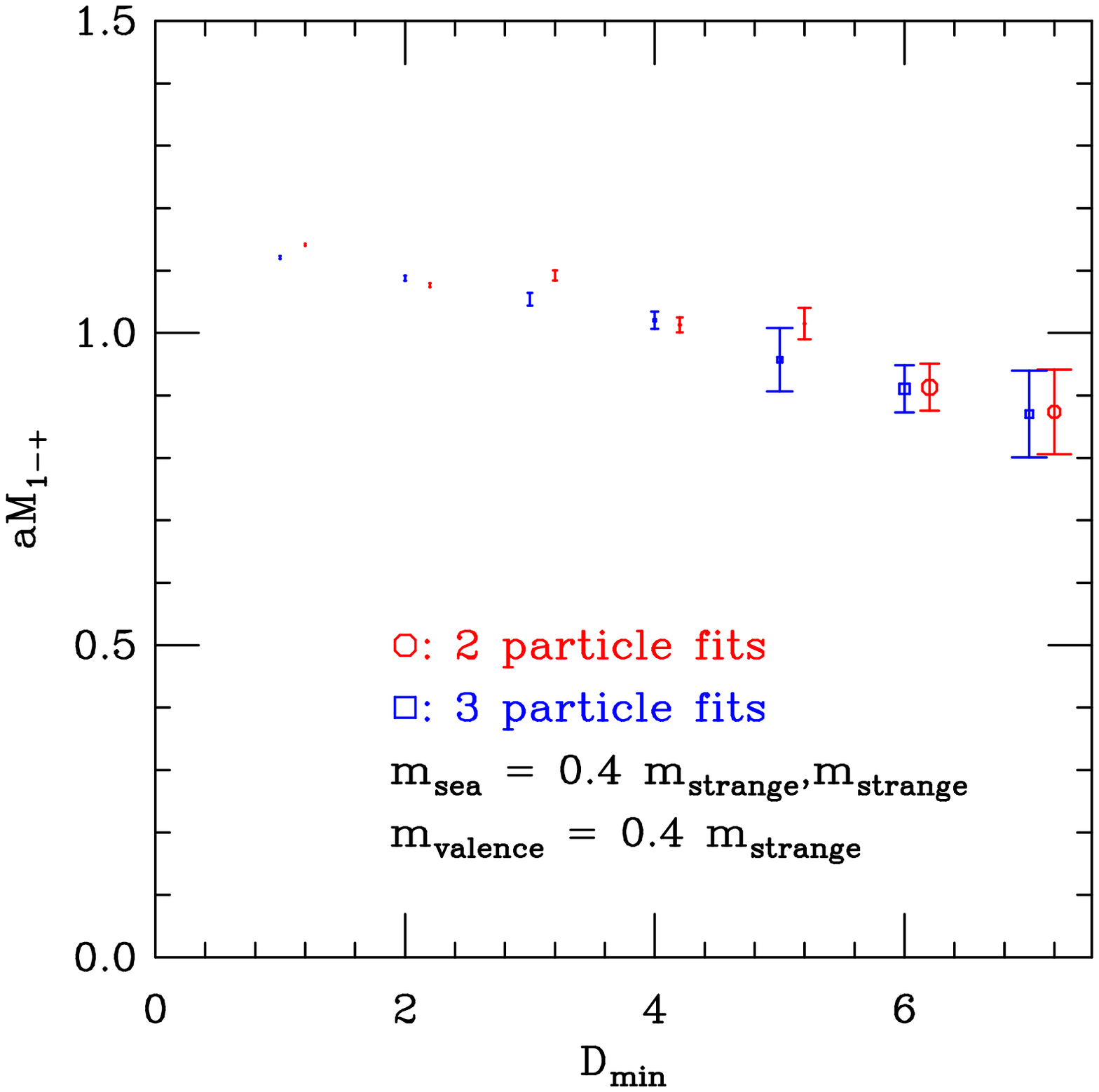}}
\caption{$aM_{1^{-+}}$ {\em vs.} $D_{\rm min}$ for $10/g^2=7.11$ with
three dynamical quarks with masses $am_{\rm light}=0.0124$
and $am_{\rm heavy}=0.031$.  The valence quark mass is
$am_{\rm valence}=0.031$.
Notation is the same as in Fig.~\protect\ref{fits_b840m040_fig}.
\label{fits_b711m031_fig}}
\end{figure}

\begin{figure}[t]
\resizebox{5.0in}{!}{\includegraphics{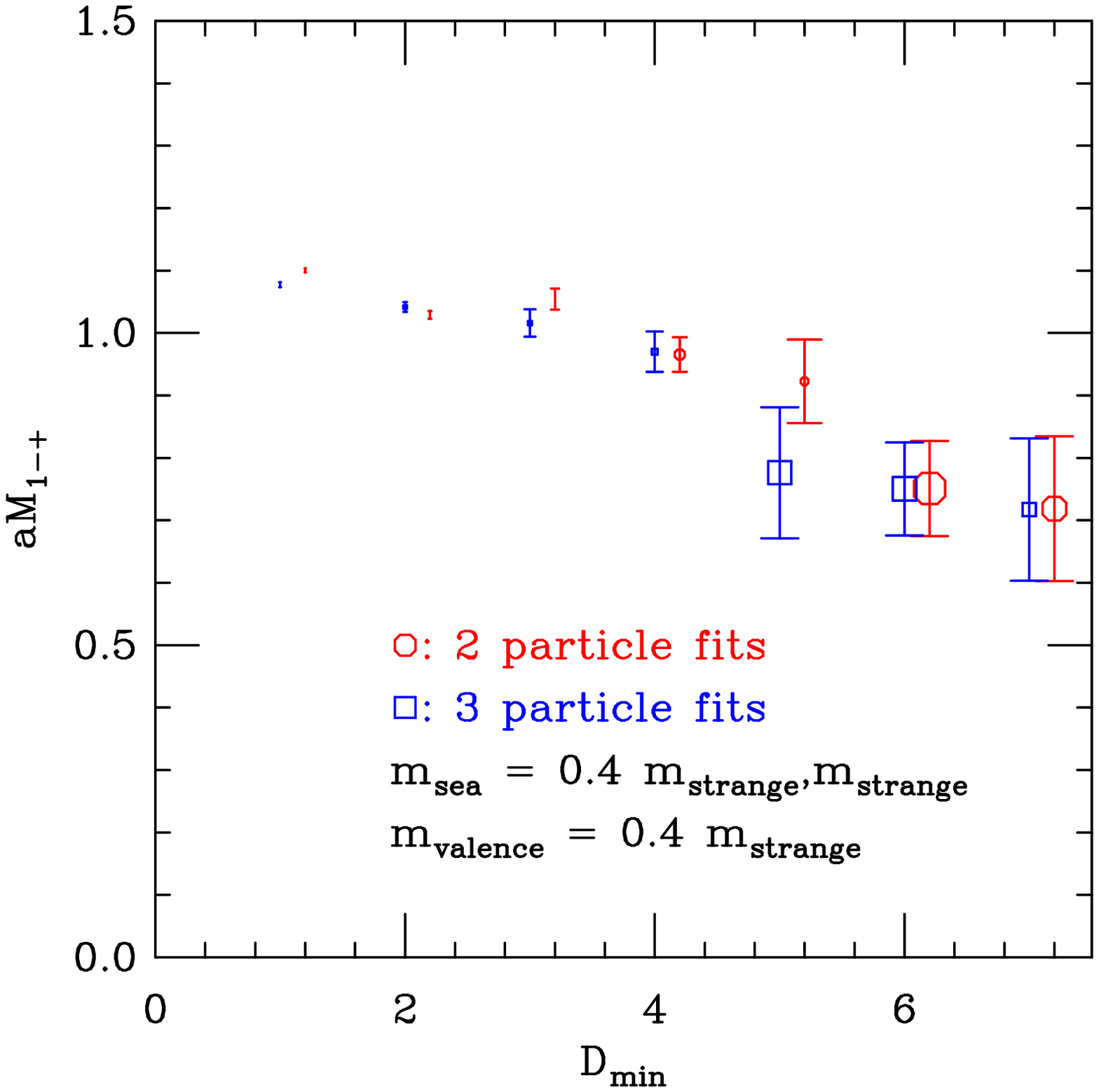}}
\caption{$aM_{1^{-+}}$ {\em vs.} $D_{\rm min}$ for $10/g^2=7.11$ with
three dynamical quarks with masses $am_{\rm light}=0.0124$
and $am_{\rm heavy}=0.031$.  The valence quark mass is
$am_{\rm valence}=0.0124$.
Notation is the same as in Fig.~\protect\ref{fits_b840m040_fig}.
\label{fits_b711m0124_fig}}
\end{figure}

The lattices with $m_{u,d}=0.4m_s$ proved more interesting and difficult. 
The $1^{-+}-1^{++}$ propagator for valence mass $am_q=0.0124$
for this ensemble is shown in Fig.~\ref{propm0124nf0b711}.
Fits to the $1^{-+}$ mass for both valence masses are
illustrated in Figs.~\ref{fits_b711m031_fig} and \ref{fits_b711m0124_fig}.
The fitted mass agrees with those of the 
quenched and three-flavor results within two standard deviations, but with 
larger systematic errors, estimated from the dependence on fit range.

In the case of the light valence quark 
($ma=0.0124$), we were unable to say much about the $1^{-+}$ hybrid mass
with any confidence.  It is apparent from visual examination of the propagator
(Fig. \ref{propm0124nf0b711}) that there is a lessening of the overall slope,
suggesting that the non-oscillating piece may not be consistent with a single
exponential. Indeed, the fits were very range dependent. Together these
factors indicate the presence of lighter $1^{-+}$ states, likely to be
the states of two mesons into which the hybrid can decay.
However, with the statistics available to us, we are unable to get convincing plateaus
in the fits with more than one exponential in the $1^{-+}$ channel.

We performed a linear extrapolation in quark mass of the quenched results 
to the physical value of  $(m_{\rm PS}/m_V)^2$. Because the calculations at
the two quark masses were done on the same set of quenched configurations, 
they are highly correlated, and a single elimination jackknife method was
used to estimate the statistical error of the extrapolation.

\begin{table}
\begin{tabular}{lclllc|cll}
$10/g^2$ & $m_{\rm sea}a$ & $m_{\rm val}a$ 
& $a^2\sigma$ & $r_1/a$& $N_{\rm configs}$ &  Range & $aM_{1^{-+}}$ & c.l.\\
\hline
8.40  & ---          & 0.040  & 0.0499(5) & 3.730(7)  & 416
    & 4--15 & 1.062(12)(20)& 0.27 \\
8.40  & ---          & 0.016  & 0.0499(5) & 3.730(7)  & 416
    & 4--15 & 0.973(26)(20) & 0.49\\
\hline
7.18  & 0.031 & 0.031  & 0.0405(7) & 3.829(13)  & 509
    & 5--15 & 0.986(30)(30)&0.83\\
7.11  & 0.0124, 0.031 & 0.031  & 0.0424(9)  & 3.708(14) & 526
    & 6--15 & 0.911(34)(100) & 0.25 \\
7.11  & 0.0124, 0.031 & 0.0124 & 0.0424(9) & 3.708(14) & 526
    & na & & \\

\end{tabular}
\caption{\label{hybrids_table}Summary of hybrid meson simulation parameters and results.
All lattices have dimensions $ 28^3 \times 96 $.
The $1^{-+}$ (hybrid) mass fits are all three particle fits.
The second error on the hybrid mass estimates is an estimate of the
possible systematic error from our choice of fit range.
}
\end{table}

\begin{table}
\begin{tabular}{lclllll}
$10/g^2$ & $m_{\rm sea}a$ & $m_{\rm val}a$ 
& $aM_{\rm PS}$ & $aM_{\rm V}$ & $aM_N$ & $aM_{dec}$ \\
\hline
8.40  & ---          & 0.040  
    & 0.348 & 0.523(3) & 0.771(2) & 0.855(17) \\
8.40  & ---          & 0.016  
    &0.223 &0.468(3) & 0.633(2) & 0.749(18) \\
\hline
7.18  & 0.031 & 0.031  
    &0.320 & 0.478(1) & 0.699(1) & 0.766(2) \\
7.11  & 0.0124, 0.031 & 0.031  
    & 0.326 &  0.479(2) & 0.710(2) & na \\
7.11  & 0.0124, 0.031 & 0.0124 
    & 0.206& 0.414(2) & 0.579(3) & 0.692(4) \\

\end{tabular}
\caption{\label{conventional_table}Preliminary values for conventional hadron masses
at the hybrid mass simulation parameters.
Statistical errors on the 
pseudo-scalar meson mass, $aM_{\rm PS}$ are smaller than the precision shown.
Pseudo-scalar and vector meson masses for the $10/g^2=8.4$ quenched points were obtained
from interpolation or extrapolation from results at valence masses 0.015 and 0.030.
}
\end{table}

\section{DISCUSSION AND CONCLUSIONS}

There are several sources of systematic error to be estimated.  The largest
of these, namely use of the quenched approximation, is inextricably mixed
with the problem of determining the overall scale, or lattice spacing, so
we will discuss these issues together.

The first source of systematic error is due to the
possibility of mixing of higher mass states in the $1^{-+}$ propagators.
As described above we estimate this by looking at the 
mass range one might get by a reasonable variation of the 
fitting parameters.

\begin{figure}[t]
\resizebox{5.0in}{!}{\includegraphics{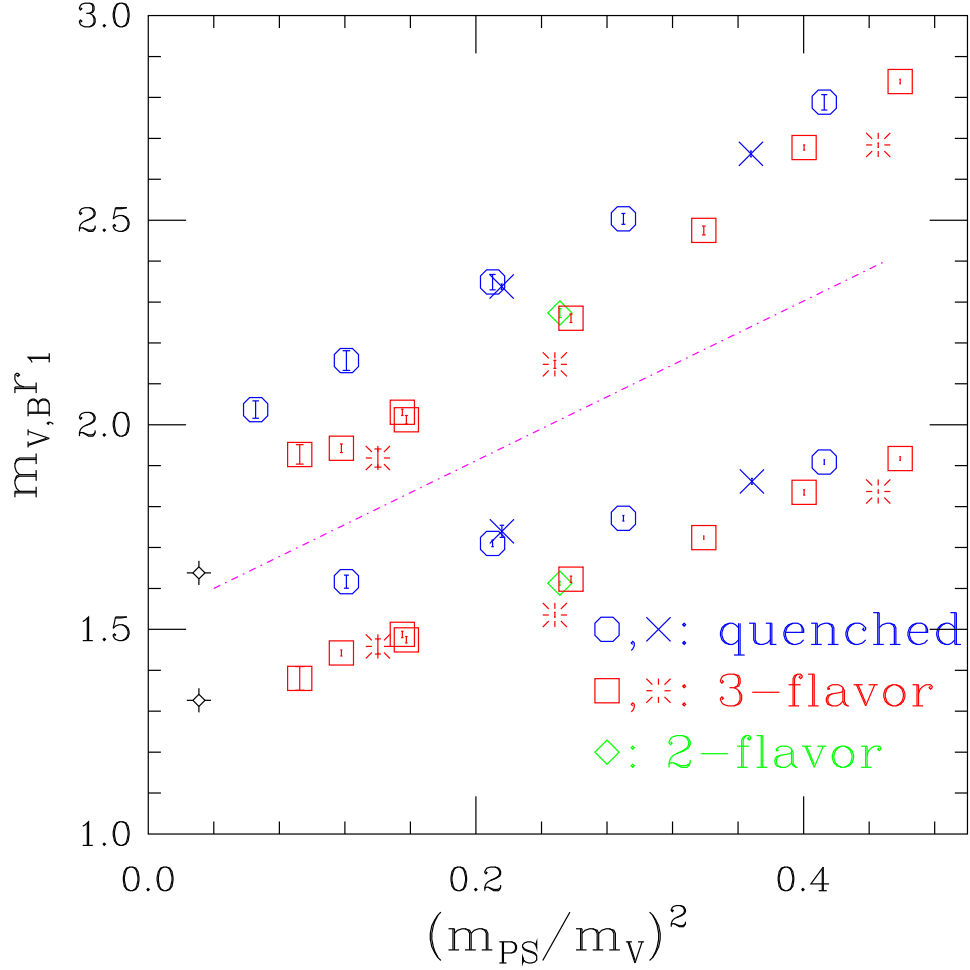}}
\caption{Vector meson (``$V$'')  and octet baryon  (``$B$'') masses
in units of $r_1$, which is
defined from the static quark potential by $r_1^2 F(r_1) = 1.0$.
This graph contains points from quenched simulations with 
$a \approx 0.13$ fm (\tmpblue octagons) and 0.09 fm (\tmpblue crosses),
and from simulations with three flavors of dynamical quarks (two light and
one strange quark) at $a \approx 0.13$ fm (\tmpred squares) and
0.09 fm (\tmpred bursts).  The \tmpgreen diamond is from a two
flavor simulation with $a \approx 0.13$ fm.
Points above the dashed line are baryon masses, and those below the
dashed line vector meson masses.
\label{mhadron_r1_fig}}
\end{figure}

\begin{figure}[t]
\resizebox{5.0in}{!}{\includegraphics{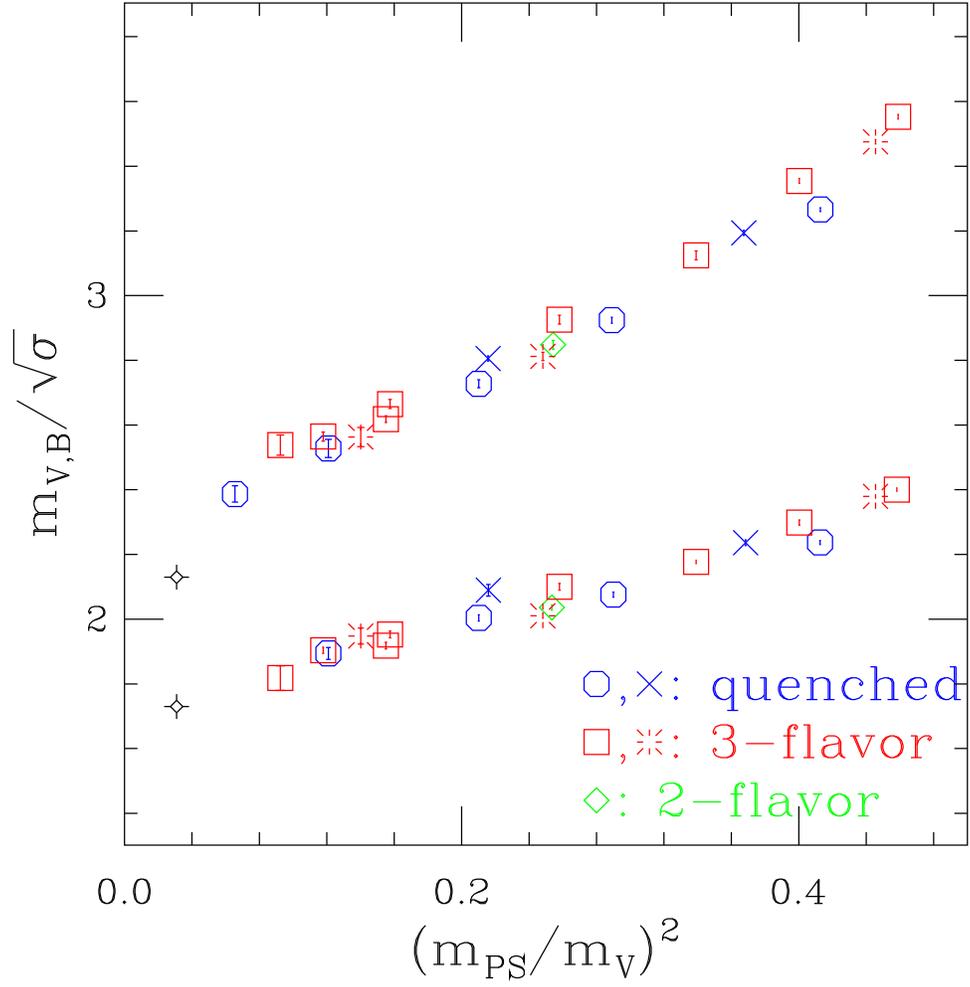}}
\caption{
Vector meson (``$V$'')  and octet baryon  (``$B$'') masses
in units of the square root
of the string tension.
The meaning of the symbols is the same as in Fig.~\protect\ref{mhadron_r1_fig}
\label{mhadron_sigma_fig}}
\end{figure}

We also have effects of finite lattice spacing.
We obtained these results on lattices with $a\sim 0.09$ fm.
For the conventional hadrons, we have masses at both $a\approx 0.13$ fm
and $a\sim 0.09$ fm (Figs.~\ref{mhadron_r1_fig} and \ref{mhadron_sigma_fig}).
Since errors with this action are expected to be order $a^2 g^2$, and
the finer lattice spacing is about $1/\sqrt{2}$ times the coarser lattice spacing,
we expect that the difference between $a\sim 0.09$ fm and $a=0$ masses
is comparable to or slightly smaller than the difference between $a\sim 0.13$
and $0.09$ fm.  For the quenched $m_\rho/\sqrt{\sigma}$ and $m_N/\sqrt{\sigma}$ 
we see differences as large as 3\% between the two lattice spacings, and
a difference of about 2\% in the ratio $m_N/m_\rho$ at the light quark
mass.  Differences are smaller at the heavier mass --- less than 1\% in the
nucleon to rho mass ratio.
Therefore we expect effects of finite lattice spacing on our results 
based on hadron mass ratios to
be around 1\% for strange quarks, and we will use an estimate of 3\%
for light quarks.

The finite size of the entire lattice also introduces systematic error.
The $28^3\times 96$ lattice corresponds to a box $(2.5 {\rm\ fm})^3 \times 8.6$ fm.
In one case, three flavor QCD with light quark mass about 0.2 times the
strange quark mass with a lattice spacing of 0.13 fm,
we have calculated light hadron masses both in a
$2.5$ fm box and on a larger $3.6$ fm spatial lattice.
The $\rho$, $\phi$ and nucleon masses decrease by a barely significant
0.9(7)\%, 0.25(25)\% and 0.9(6)\% respectively as the lattice size goes
from 2.5 to 3.6 fm.
Since these effects are expected to fall exponentially with lattice size, 
we can simply take these numbers as an estimate of the effect of the 2.5 fm
box size on the light hadron masses.
However, hybrids are expected to be rather 
extended objects and may feel the influence of a finite lattice more than 
smaller particles, so we will use an estimate of 2\% for this systematic error.

The largest systematic errors come from use of the quenched approximation,
from the choice of quantity used to set the lattice scale, and the necessity
for an extrapolation to the physical value of the valence quark mass.
These effects are interrelated and so must be discussed together.

\begin{figure}[t]
\resizebox{6.0in}{!}{\includegraphics{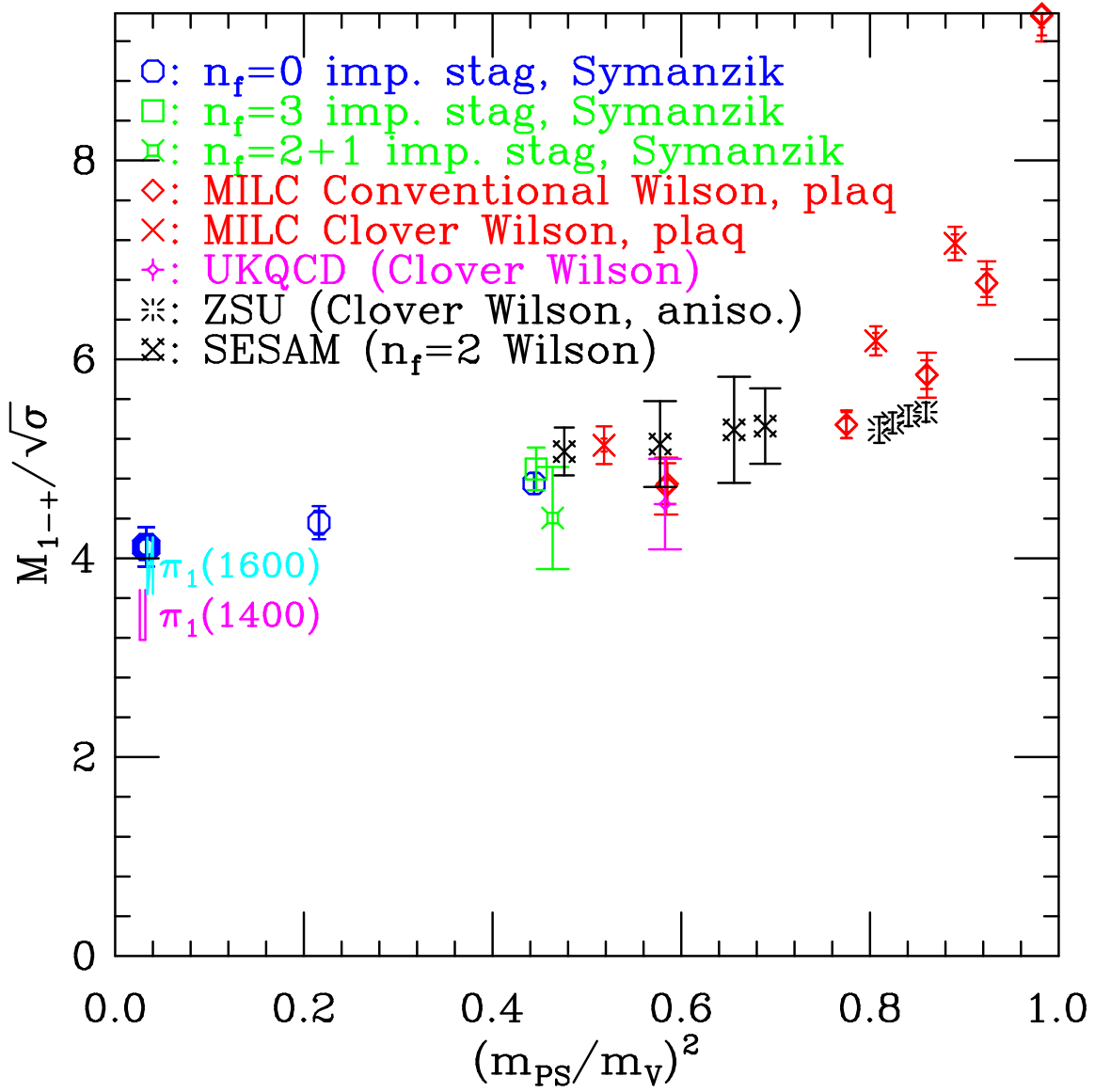}}
\caption{\label{summary_sigma}
Summary of $1^{-+}$ hybrid meson mass predictions as a function
of $(m_{\rm PS}/m_V)^2$. The bold octagon represents the linear 
extrapolation of $n_f=0$ data to $(m_{\rm PS}/m_V)^2=0.033$.
The improved staggered points are from this work, while the
earlier data is from Refs~\protect\cite{Bernard:1997ib,Lacock:1996ny,ZSU,Lacock_dyn}
}
\end{figure}

The hybrid mass estimates obtained above are in units of the inverse lattice
spacing $a^{-1}$, so to convert these to physical units we need to know $a$.
The lattice spacing is determined by calculating some quantity that is known
from experiment.  In other words, the simulation actually produces the ratio
of the hybrid mass to some other dimensionful quantity.  In a simulation with
sea quark masses at their physical values, the choice of quantity to fix the
lattice spacing would be just a question of convenience.  However, in the
quenched approximation, we will not get the real world values for ratios
of masses, so there is an important choice to be made.
Because it is easily measured, and because it does not require an extrapolation
in valence quark masses, the static quark potential is often used to
determine the lattice spacing.
In particular, we may use the string tension,
$\sqrt\sigma \approx 440\ {\rm MeV}$,
the coefficient of the linear term in $V(r)$.
We might also use $r_0 \approx 0.50$ fm or $r_1 \approx 0.34$ fm, which are
defined by $r_x^2 F(r_x) = 1.65$ or $1.00$ respectively.
However, the shape of the static quark potential in quenched QCD differs from
the shape with three dynamical flavors~\cite{MILC_spectrum}.
Hybrid mesons are expected to be large
hadrons where the quarks are more likely to be in the 
linear part of the static quark potential, where $\sigma$ is defined, 
rather than the region of crossover between Coulombic and linear
behavior, where $r_0$ and $r_1$ are defined.
This suggests that
plotting results in units of the string tension might minimize
(although by no means eliminate!) effects of quenching.
This expectation is borne out by calculations of the conventional
hadron spectrum with this same improved action, where using $\sigma$
to define the lattice spacing produces better agreement of the quenched
and three flavor results than using $r_1$~\cite{MILC_spectrum}.
Figures~\ref{mhadron_r1_fig} and \ref{mhadron_sigma_fig} illustrate this 
with rho and nucleon masses plotted in units of $r_1^{-1}$ and 
$\sqrt{\sigma}$ respectively.
Since one of our important goals is to compare quenched and three
flavor results, we therefore plot our results in units of the string
tension.  We also wish to compare our results with earlier results, and
for this purpose the string tension in other published simulations
is either available or can be reasonably estimated.
In Fig. \ref{summary_sigma} we summarize our results  along with the results 
of previous Wilson quark studies by the MILC collaboration ~\cite{Bernard:1997ib}, 
the UKQCD collaboration
~\cite{Lacock:1996ny}, the SESAM collaboration ~\cite{Lacock_dyn},
 as well as recent results from the Zhongshan University group~\cite{ZSU}
using Wilson quarks on an anisotropic lattice.
We use the string tension $\sigma$ to establish the lattice length scale 
and plot $M_{1^{-+}}/\sqrt{\sigma}$.
Our results are consistent with the earlier results at heavier quark masses.

To compare with experiment, we need to convert $M_H/\sqrt{\sigma}$
to physical units.
Unfortunately, although phenomenological estimates are available, the
string tension is not a parameter that is well known from experiment.
The obvious workaround is to determine the string tension from the
lattice results for $m_\rho/\sqrt{\sigma}$ etc., which in the end
means that we are using the light hadron spectrum to set the length
scale.  Since ratios of quenched hadron masses are not quite those of
the real world, we will get different estimates of the length scale
depending on which hadron we choose.
For the $s \bar s$ hybrid, the most reasonable choice for setting
the length scale is a hadron with valence quark masses
at the same value --- the $\phi$ meson or $\Omega^-$ baryon,
which means that we
are essentially quoting $M_H/M_\phi$ or $M_H/M_{\Omega^-}$ with the
quenched $\phi$ mass and $\Omega^-$ masses
defined to be 1020 MeV and 1672 MeV.
Estimating the masses of the conventional hadrons on our quenched lattices
from a linear extrapolation of results at $am_q=0.015$ and $0.030$,
and setting the quenched string tension from the $\phi$ or $\Omega^-$ gives
$\sqrt{\sigma} = 436(4)$ or $437(9)$ MeV respectively. (This remarkable agreement is
surely coincidence, since other hadron mass ratios on these lattices
differ by much larger amounts from the real world.)
To estimate the light quark hybrid mass in MeV, we might use
these estimates of $\sqrt{\sigma}$, or equally well argue that
we should use light quark hadrons for comparison.
Using the linearly extrapolated or interpolated $\rho$, $K^*$,
$N$ or $\Delta$ masses to set the scale gives quenched 
$\sqrt{\sigma}$ of 389(5), 410(4), 380(5) or 400(21) MeV respectively,
showing statistical errors only.
These estimates are in reasonable agreement with phenomenological estimates
from potential models on charmonium and bottomonium spectroscopy; for
example $\sqrt{\sigma} = 384$ MeV or $427$ MeV in Refs.~\cite{RICHARDSON}
and ~\cite{EICHTEN} respectively.
Thus in estimating light quark hybrid
masses in MeV we might consider a range of possible
values for the quenched $\sqrt{\sigma}$ from around 380 to 440 MeV.

We begin with estimates for the $s\bar s$ hybrid masses.
As mentioned above, it seems most consistent to use masses of hadrons
made from strange quarks to set the lattice spacing in this case.
If we use the $\phi$ meson to set the length scale, using the results
in Tables \ref{hybrids_table} and \ref{conventional_table} we find, with
statistical error only
\BE M_{H,s \bar s} = 1020 \ {\rm MeV}\  \LP\frac{1.062(12)}{0.523(3)}\RP = 2071(26) {\rm \ MeV} \ \ \ .
\EE
Systematic errors include fit choice, nonzero lattice spacing, finite
spatial size, and effects of quenching.  The first three have
been discussed above.  Effects of quenching can be estimated in
part from the variation of our mass estimates among different ways
of fixing the lattice scale, and in part from differences of other
hadronic ratios between full and quenched QCD, as for example in Fig.~\ref{mhadron_sigma_fig}.
Here we include what we expect is a fairly conservative 5\% error for
this effect, thus estimating
\BEA M_{H,s \bar s} &=& 2071(26)(39)(1\%)(2\%)(5\%) \\
&=& 2071(120) {\rm \ MeV}\ \ \ , 
\EEA
where the errors are statistical, fit choice, lattice spacing, box size and
quenching respectively.
A similar calculation using the $\Omega^-$ mass to set the scale gives
\BEA 
M_{H,s \bar s} &=& 1672 \LP\frac{1.062(12)}{0.855(17)}\RP \nonumber\\
&=& 2077(48)(39)(1\%)(2\%)(5\%)\nonumber\\ 
&=& 2077(129) {\rm \ MeV}\ \ \ . 
\EEA
We might also use the mass of a fictional octet baryon made from three
quarks with the mass of the strange quark, assigning it a mass of
$M_{sss} = m_N + \frac{3}{2} \LP m_\Xi - m_N \RP = 1507\ {\rm MeV} $:
\BEA 
M_{H,s \bar s} &=& 1507 \LP\frac{1.062(12)}{0.771(2)}\RP\nonumber \\
&=& 2075(24)(39)(1\%)(2\%)(5\%)\nonumber\\
 &=& 2075(119) {\rm \ MeV}\ \ \ .
\EEA
These three estimates are in remarkably close, and doubtless partly fortuitous, agreement.

Repeating this calculation with the three flavor lattices with $m_{u,d}=m_s$
with the $\phi$, $\Omega^-$ and $sss$ baryon setting the scale produces
\BEA
M_{H,s \bar s} &=& 1020 \LP\frac{0.986(30)}{0.4778(9)}\RP\nonumber \\
 &=& 2105(64)(64)(1\%)(2\%)(3\%)\nonumber \\
 &=& 2105(120) {\rm \ MeV} \\
M_{H,s \bar s} &=& 1672 \LP\frac{0.986(30)}{0.7659(24)}\RP\nonumber\\
 &=& 2152(66)(66)(1\%)(2\%)(3\%)\nonumber\\
 &=& 2152(123) {\rm \ MeV} \\
M_{H,s \bar s} &=& 1507 \LP\frac{0.986(30)}{0.6991(10)}\RP \nonumber\\ 
 &=& 2125(65)(65)(1\%)(2\%)(3\%)\nonumber\\ 
 &=& 2125(121) {\rm \ MeV} 
\EEA
respectively.
Here we have assigned an error of 3\% for the partial quenching, or the remaining extrapolation
of the sea quark masses to their physical values.
Finally, we made an estimate of the $s\bar s$ hybrid mass from the run with
$m_{u,d} =0.4 m_s$.  Although the error on this estimate, mostly coming
from the choice of fit range, is too large for it
to be very useful, we include it for completeness.
\BEA
M_{H,s \bar s} &=& 1020 \LP\frac{0.911(34)}{0.4792(16)}\RP\nonumber\\
 &=& 1939(73)(213)(1\%)(2\%)(2\%)\nonumber\\
 &=& 1939(233) {\rm \ MeV} \ \ .
\EEA
Since the sea quarks here are much lighter, we used 2\% as our estimate of the 
systematic error from partial quenching in this number.
We can summarize this with an estimate of $2100 \pm 120$ MeV for the mass
of the $s \bar s$ $1^{-+}$ hybrid meson.

To estimate the mass of a light quark $1^{-+}$ hybrid meson we use the jackknife
extrapolation of the quenched results to $\LP m_{PS}/m_{V}\RP^2 = 0.033$, $am_H = 0.919(39)$.
If we use the $\phi$ to set the scale, this would correspond to a mass of
\BEA
M_{H,u \bar u} &=& 1020 \LP\frac{0.919(39)}{0.523(3)}\RP\nonumber\\
 &=& 1792(77)(36)(3\%)(2\%)(5\%)\nonumber\\
 &=& 1792(139) {\rm \ MeV} \ \ \ ,
\EEA
with similar results using the $\Omega^-$ or $sss$ baryon.
However, if we were to use the smaller estimates of the string tension
obtained from linear extrapolations of light quark hadron masses to the
physical light quark mass, we would obtain smaller values around 1600 MeV.
As discussed above  we have assigned a larger 3\% systematic
error for the effect of nonzero lattice spacing.
We have also assigned a larger 5\% error from quenching and chiral extrapolation.
One reason that a larger systematic error is required here is that we
are estimating the lattice spacing in large part from hadrons made up of
strange quarks.  Our strange quark mass was fixed by tuning the pseudo-scalar
to vector meson mass ratio, and would have come out slightly different
if we had used some other quantity.  The effect of uncertainty
in fixing the strange quark mass mostly cancels from mass
ratios of hadrons made up of strange quarks, such as $M_{H,s \bar s} / M_\phi$,
but will be present when quantities such as $M_\phi$ are used in estimating
the mass of light quark hadrons.
More evidence that this larger systematic error is required is seen in the extrapolations
of conventional hadron masses to the physical light quark mass.   
If a naive linear extrapolation is made, and the resulting masses used to
set the scale for the $1^{-+}$ hybrid mass, the close agreements of the scales
from various conventional hadrons that we found when using hadrons made from
strange quarks is no longer present, as seen in the string tension estimates
above.

Given the systematic errors from quenching and chiral extrapolation,
our estimate for the mass of the light quark $1^{-+}$ meson is not
inconsistent with the experimental candidate $\pi_1(1600)$.
In Fig.~\ref{summary_sigma} we include the 
$1^{-+}$ experimental candidates $\pi_1(1400)$ and $\pi_1(1600)$ 
at the physical value of $(m_{\rm PS}/m_V)^2=(m_{\pi}/m_\rho)^2=0.033$.
These particles are represented by \tmpmagenta \tmpand \tmpcyan vertical
bars, offset slightly to the left or right for clarity, representing
the range of values for the quenched string tension
from 380 to 440 MeV.

\begin{figure}[t]
\resizebox{6.0in}{!}{\includegraphics{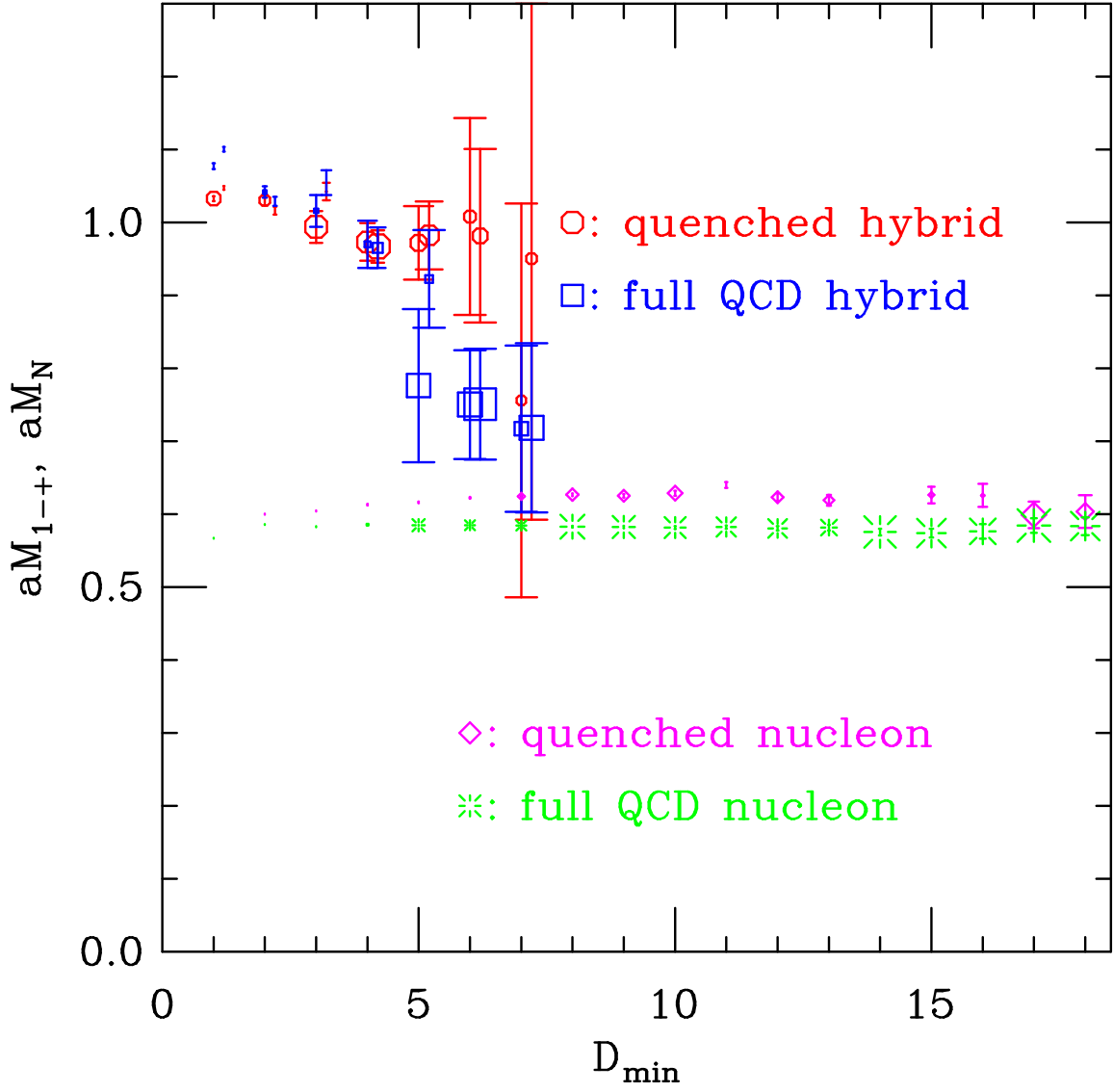}}
\caption{\label{fits_hyb_nuc} Hybrid and nucleon mass
fits in quenched and full QCD with
light dynamical quark mass $am_{\rm light} \approx 0.4 m_s$.
The valence quark mass is about $0.4 m_s$, which is $am_{\rm valence}
=0.016$ for the quenched case and $0.0124$ for the three
flavor case.
}
\end{figure}

The $m_{u,d}=0.4m_s$ data illustrates that dynamical quarks introduce 
new and significant processes that contribute to the $1^{-+}$ propagator. 
On this same set of lattices, mass fits for stable hadrons, even with 
$ma=0.0124$
valence quarks, display plateaus as functions of minimum included
distance, $D_{\rm min}$ with fixed maximum distance, $D_{\rm max}$.
The plateaus are similar for quenched and full QCD.  
In contrast, for the $1^{-+}$, the full QCD fits do not show even the 
shorter plateau found in the quenched fits.
We illustrate this by comparing fit plots for quenched and full QCD hybrids 
and nucleons in Figure \ref{fits_hyb_nuc}.
Fit plots for nucleon and quenched hybrids show a plateau,
indicating the propagator has a single exponential form in the region 
$D_{\rm min}$ to $D_{\rm max}$. The full QCD hybrid fit plot deviates from 
a plateau in a significant manner --- at minimum distance five, in the 
range which
we have generally used for our quoted mass estimate, the low mass
full QCD fits drop to a smaller value.
Though quenching often introduces a systematic effect in the mass, this 
propagator is different in a way that suggests
mixing of more than one exponential,
representing  propagators of different states with $J^{PC}=1^{-+}$.
Our hybrid propagators with light, dynamical quarks show features that are 
not evident either in hybrid propagators with heavier or quenched quarks, or
in stable hadron propagators even with light dynamical quarks. 


Four-quark states, molecular states of two mesons, or two independent
mesons
can have $J^{PC}=1^{-+}$ without the gluonic excitations. For example the
combination of $b_1+\pi$ can give $1^{-+}$ with $I=1$, and as the sum of these masses
is less than the predicted mass of the lowest $1^{-+}$ hybrid, we
expect that dynamical quarks introduce the possibility of the hybrid decaying 
into this two-meson state.
In fact, at the values of the quark masses that we used
the $1^{-+}$ energies found in our dynamical simulations, while similar
to the quenched hybrid masses, are also very close to the expected decay channel
masses.
For the run with three degenerate sea quarks at $m_s$, our $1^{-+}$
mass is $am_H=0.97(3)(3)$, very close to the sum of the ``$\pi$''and ``$b_1$''
masses:  $am_\pi+am_{b1} = 0.32+0.68=1.00$.
For the run with $m_{u,d}=0.4m_s$, we would expect decays into a pseudo-scalar $K$
and a P-wave strange meson - a $K_1$.  Again, our estimated mass for the $s \bar s$
$1^{-+}$, $am_H=0.90(4)(10)$,  is close to the sum $am_K + am_{K1} = 0.27+0.63 = 0.90$.

We now have ahead of us the task of 
understanding these contributions so that we 
can make useful predictions of the $1^{-+}$ hybrid mass in the presence of 
dynamical quarks.
It is clear from our results with dynamical quarks that it will not be
sufficient to simply do the same analysis that was done on the
quenched gauge configurations, simply replacing them
with full QCD configurations.
One obvious avenue that may shed some light is to measure 
cross-correlators  between the $\rho \times B$ operator and the two-meson
state, as was explored with Wilson quarks in Ref.~\cite{Bernard:1997ib}.
A more detailed study along these lines in the static quark (heavy quark)
limit has been done by the UKQCD collaboration~\cite{UKQCDDECAY}.
It may also be useful to study the dependence of the exotic energy as
a function of valence quark mass (possibly with fixed sea quark mass)
to look for an avoided level crossing as the decay threshold is crossed,
as was done for the non-exotic $0^{++}$ meson in Ref.~\cite{MILC_spectrum}.

\section*{ACKNOWLEDGEMENTS}
Computations for this work were performed at the San Diego Supercomputer
Center (SDSC), the Pittsburgh Supercomputer Center (PSC), Oak Ridge National Laboratory (ORNL)
and the National Energy Resources Supercomputer Center (NERSC).
This work was supported by the U.S. Department of Energy under contracts
DOE -- DE-FG02-91ER-40628,      
DOE -- DE-FG02-91ER-40661,      
DOE -- DE-FG02-97ER-41022       
and
DOE -- DE-FG03-95ER-40906       
and National Science Foundation grants
NSF -- PHY99-70701              
and
NSF -- PHY00--98395.            


\end{document}